\documentclass[11pt]{article}
\usepackage{fleqn,cospar}

\usepackage{url}

\usepackage{graphicx}
\usepackage[figuresright]{rotating}

\def\asca{{\it ASCA}}

\def\xte{{\it RXTE}}

\title{X-RAY VARIABILITY OF AGN AND CORRELATIONS WITH SPECTRAL PROPERTIES}

\author{K. Nandra\address{Mail Code 662, NASA/GSFC, Greennelt,
MD 20771, USA}\address{Universities Space Research Association}}

\begin{document}

\maketitle

\begin{abstract}
Rapid flux changes in the X-ray emission from active galactic nuclei
are commonly observed. The power-density spectra show a pseudo
power-law form with a turnover at low frequencies and a high frequency
break, similar to galactic black-hole candidates. There have been a
few claims for periodicities but these are not a well-established
property of the class. The amplitude of variability depends on a
number of source parameters, including luminosity and spectral
properties. The variability amplitude is correlated with X-ray
spectral index, and anti-correlated with the width of the permitted
optical lines. In one particularly well-observed case, NGC 7469, we
see a relationship between the X-ray and UV variability, which
indicates that the dominant emission process for the X-rays is thermal
Comptonization. Variations in the X-ray emission are related to
changes in the UV seed photons, but it appears there must also be a
mechanism - perhaps that which heats the Comptonizing corona - that
induces rapid variability intrinsic to the X-rays.
\end{abstract}

\section*{INTRODUCTION}

Active Galactic Nuclei (AGN)
\footnote{Here we restrict our discussion to radio quiet AGN, which
make up the majority of the population. Radio-loud AGN, in particular
jet-dominated blazars show extreme variability, particularly in the
$\gamma$-rays, but are atypical in this regard} show large-amplitude,
rapid variability in the X-ray band, which has been taken as evidence
that the X-rays come from very close to the central black hole. The
mechanism producing these variations remains mysterious, however. The
X-rays in AGN are thought to be produced by Compton upscattering of
softer photons in a hot ``corona'' (e.g. Sunyaev \& Titarchuk
1980). Such models fit the X-ray spectrum extremely well (Haardt \&
Maraschi 1991, 1993), particularly if the corona is patchy (Haardt,
Maraschi \& Ghisellini 1994; Stern et
al. 1995) and the particle distribution has a thermal form (Zdziarski
et al. 1994; Madejski et al. 1995; Gondek et al. 1996). The coronal
heating mechanism is not known, but some possibilities are that a
portion of the accretion flow is intrinsically hot (e.g. Shapiro,
Lightman \& Eardley 1976; Narayan \& Yi 1994) or that magnetic
reconnection creates hot flaring regions above an accretion disk
(e.g. Nayakshin \& Melia 1997; Poutanen \& Fabian 1999). In either
case, any variations in the ``seed'' photons for the Comptonizing
medium would be mimicked in the X-ray band inducing X-ray
variability. It has always appeared that the most rapid variations
occur in the X-ray band, however, which is not consistent with the
variability being merely a secondary effect, and suggesting some other
variability mechanism associated with the coronal dissipation
process.  Progress in understanding these phenomena can be made by
accumulating high quality data on the X-ray variability and comparing
these in more detail to Comptonization and other models, and to other
properties of the AGN.

\section*{X-RAY VARIABILITY CHARACTERISTICS OF AGN}

\begin{figure}[tb]
\begin{center}
\includegraphics[height=70mm]{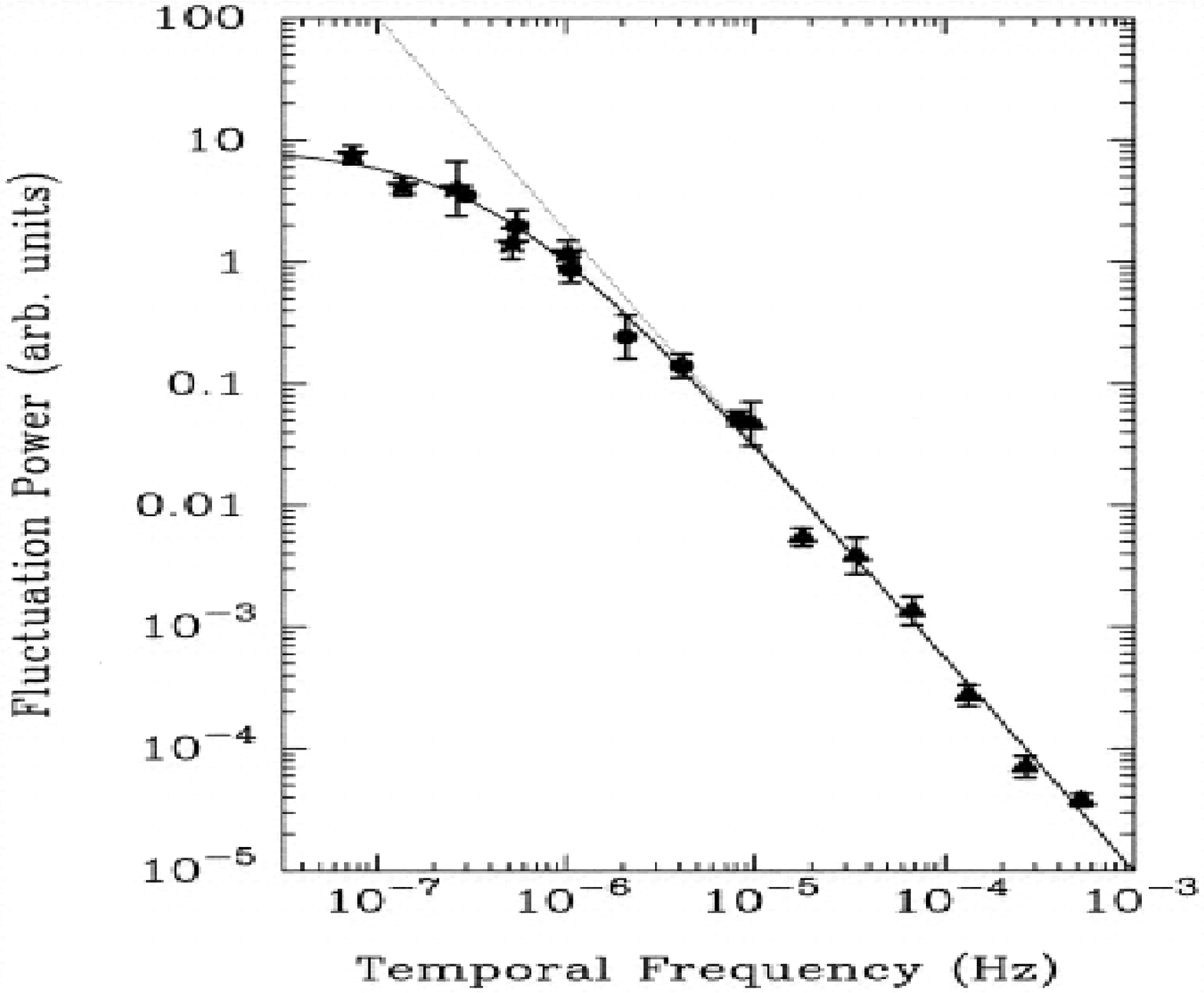}
\includegraphics[width=75mm]{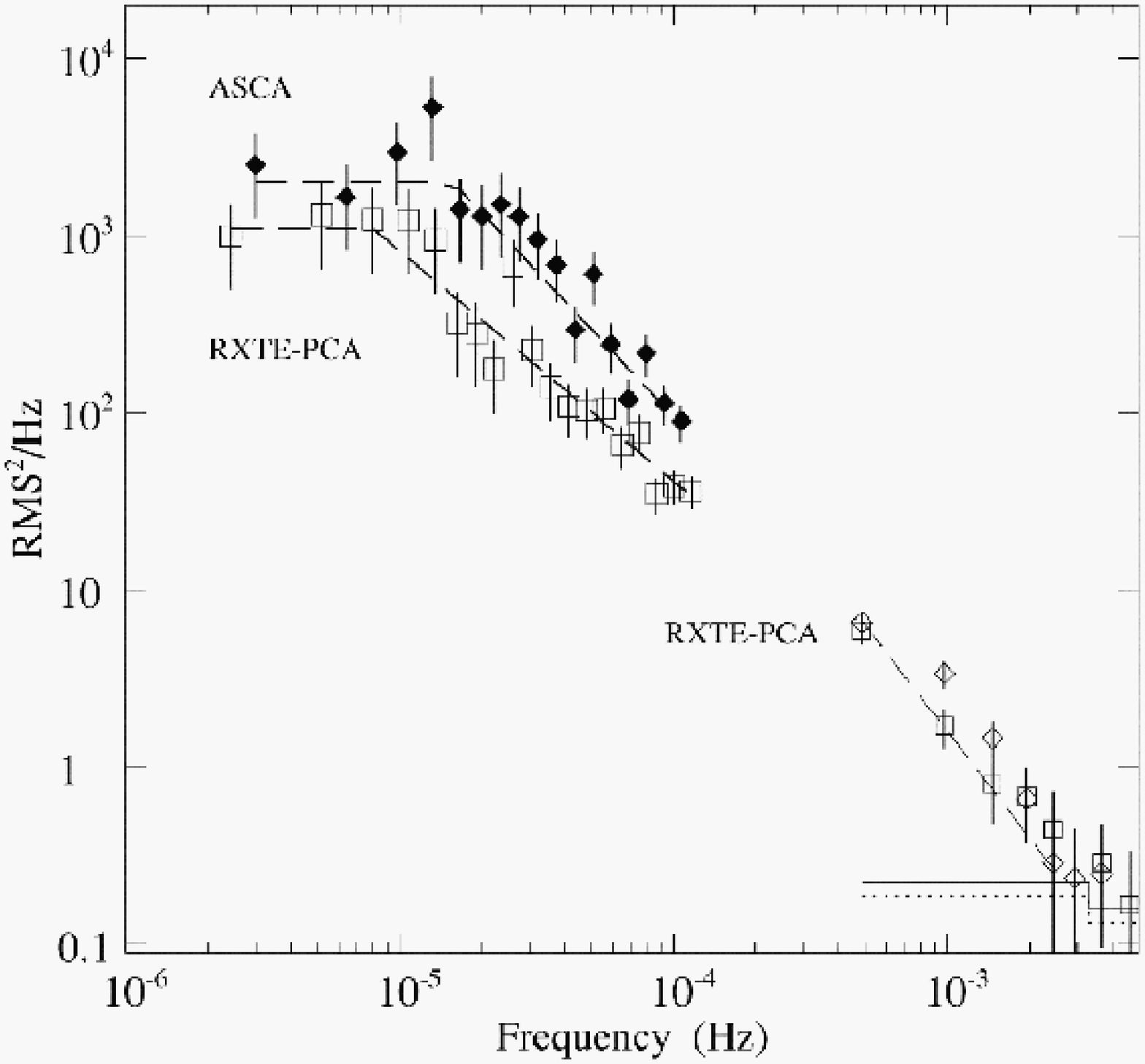}
\end{center}
\caption{(left panel) PDS of NGC 3516 obtained by the \xte\ PCA
instrument (Edelson
\& Nandra 1999). The source was sampled on time scales ranging
from minutes to years. This clearly shows a knee at low frequencies,
and exhibits a characteristic time scale of about 1 month.
(right panel) PDS of MCG-6-30-15 (Nowak \& Chiang 2000) obtained
by \xte\ and \asca. The low frequency knee is visible here also, but there
also appears to be a break at higher frequencies. Both PDS are
reminiscent of Cyg X-1. 
\label{fig:pds}
}
\end{figure}

EXOSAT demonstrated rapid variability in a number of AGN
(e.g. Lawrence et al. 1985; McHardy \& Czerny 1987) and allowed the
first definition of their Power Density Spectrum (PDS). Lawrence \&
Papadakis (1993) and Green, McHardy \& Lehto (1993) performed
systematic analyses of the EXOSAT data. Both sets of workers generally
found a featureless PDS with a steep ``red noise'' power-law
characterizing the variations. For an assumed form for the power of
$P(f)=A f^{-\alpha}$, where A is the normalization, all objects were
consistent with a single slope of $\alpha=1.5$. This steep slope
indicates that there must be a turnover (that will henceforth be
termed the ``knee'') at low frequencies, or the integral power would
be infinite. Early attempts at finding this knee (e.g. Papadakis \&
McHardy 1995) have now been improved upon using RXTE data, and there
are now several convincing reports (e.g. Fig.~\ref{fig:pds}; Table 1;
McHardy et al. 1998; Edelson \& Nandra 1999; Chiang et al. 2000). The
knee frequencies indicate characteristic time scales of orders
days-months. There has also been the report in at least one case of a
high frequency break, with the PDS of Seyfert MCG-6-30-15 showing a
further steepening above a frequency of $\sim 10^{-3}-10^{-4}$~Hz
(Fig.~\ref{fig:pds}; Nowak \& Chiang 2000).  The general form of the PDS
immediately brings to mind that of Cyg X-1 (e.g. Belloni \& Hasinger
1990), which exhibits ``white noise'' ($\alpha=0$) variability below
the knee, steepening to $\alpha=1$ and then to $\alpha=2$ above the
high frequency break.

\begin{figure}[tb]
\begin{center}
\includegraphics[width=180mm]{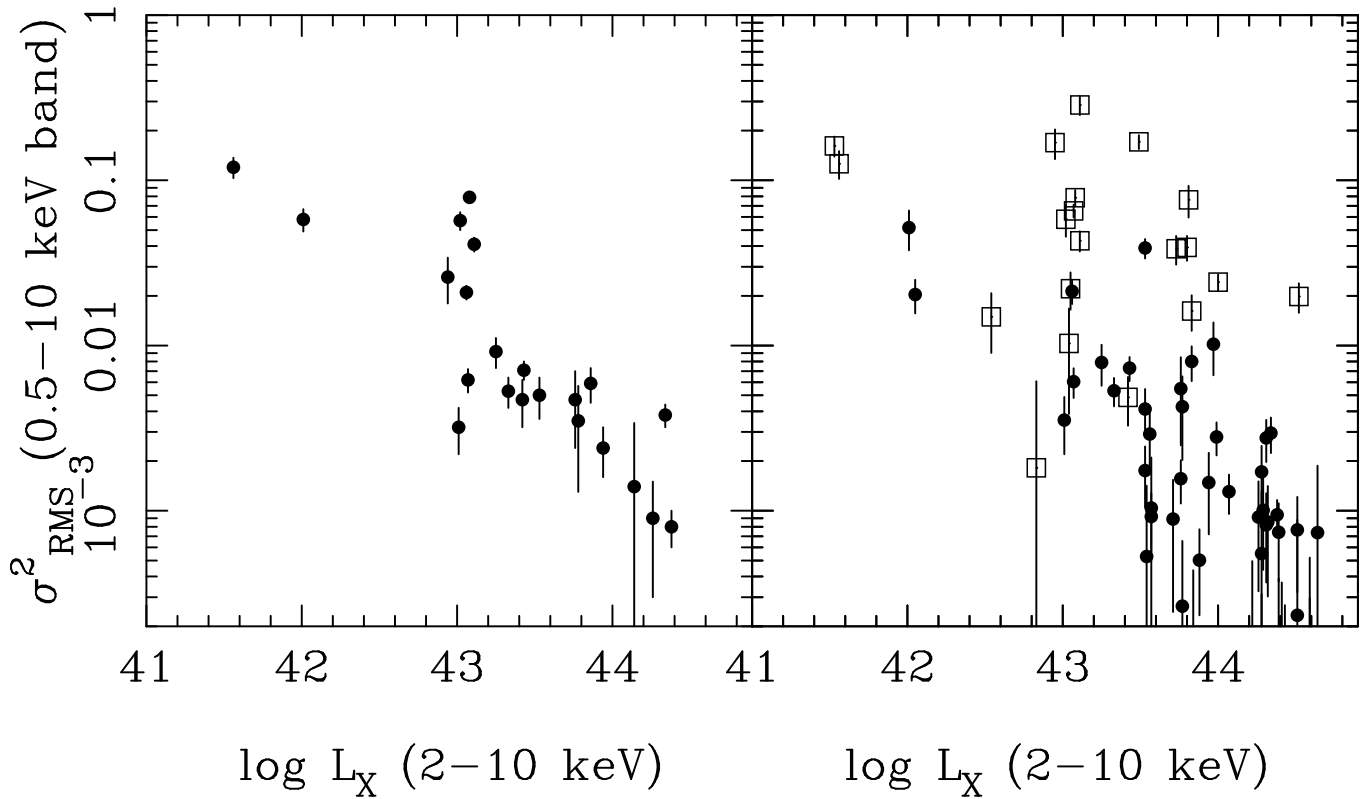}
\end{center}
\caption{Dependence of the variability amplitude (parameterized by the
``excess variance'') with luminosity for a sample of ``normal''
Seyfert galaxies (left panel; Nandra et al. 1997).  The variability
amplitude shows a strong anticorrelation. The right panel shows the
extended sample of Turner et al. (1999), which contains many more NLS1,
shown as the open squares in that panel. These tend to show a higher
variability amplitude for a given luminosity, supporting the idea that
they have higher accretion rates than ``normal'' Seyfert 1s
\label{fig:var_lum}
}
\end{figure}

Lawrence \& Papadakis (1993) first showed that the normalization of
the PDS at a fixed frequency was inversely dependent on the
luminosity. This dependence has also been confirmed by more recent
data (e.g. Fig.~\ref{fig:var_lum}; Nandra et al. 1997). This 
and the similarity with Cyg X-1 suggests
there may be a ``universal'' PDS - and variability mechanism - for
accreting black holes, which scales with some quantity like the mass
or luminosity of a given object. Table 1 shows a general trend of
decreasing break frequency with luminosity, and indeed it seems more
intuitive to associate the differences in variability characteristics
to changes in time scale, rather than in amplitude. Higher luminosity
objects presumably have higher masses, and correspondingly larger size
scales. Objects that turn over at longer time scales would tend to
show less variability at fixed frequencies above the knee.

The work of Markowitz \& Edelson (2000) has lent strong support to
this idea that the ``universal'' PDS scales in frequency, rather than
amplitude. These workers showed that the luminosity dependence of
variability amplitude is also is a function of the time scale sampled.
A shallower dependence of the variability amplitude with luminosity is
seen on long ($\sim$yr) time scales, compared to $\sim$day time
scales.  They also found less scatter in the former relation. This can
be interpreted in the framework of the universal PDS where the form
shifts along the frequency axis of the PDS plot. On long time scales,
we expect similar power in all objects, as we are sampling white noise
variability below the knee of the PDS, accounting for the shallower
relationship with luminosity. At high frequencies, above the knee,
there would be stronger relationship. Subtly different shapes above
the knee and the differing effect of the high frequency break would
cause more scatter also.

\begin{table}[tb]
\vspace{-8mm}
\begin{minipage}{120mm}
  \caption{Measurements of the low-frequency knee in accreting black holes}
\begin{tabular}{lllll}
\hline
Object 		& $\log L_{\rm X}$ & $\nu_{br}$ & Reference \\
		& (erg s$^{-1}$)   & (Hz)	& \\
\hline
\\
NGC 5548 	& 44.0 & $6 \times 10^{-8}$    & Chiang et al. (2000) \\
NGC 3516	& 43.4 & $4 \times 10^{-7}$    & Edelson \& Nandra (1999) \\
MCG-6-30-15	& 43.0 & $8 \times 10^{-6}$    & Nowak \& Chiang (2000) \\
NGC 4051	& 41.6 & $5 \times 10^{-5}$    & McHardy et al. (1998) \\
Cyg X-1		& 37.0 & $4-40 \times 10^{-2}$ & Belloni \& Hasinger (1990) \\
\hline
\end{tabular}
\end{minipage}
\hfil\hspace{\fill}
%
%
%
\end{table}%

\begin{figure}[tb]
\begin{center}
\includegraphics[width=180mm]{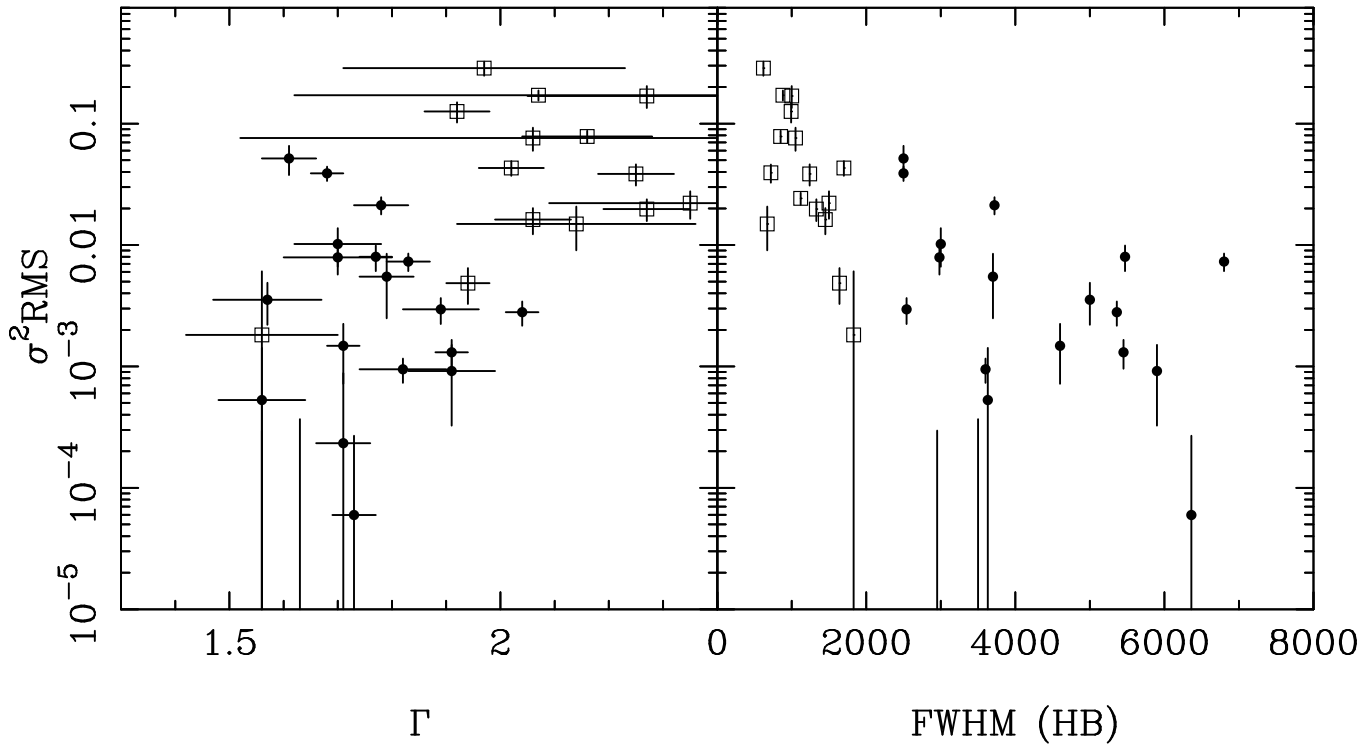}
\end{center}
\caption{Dependence of the excess variance on (left panel) the X-ray
spectral index derived in the 2-10 keV band and (right panel) the 
FWHM of the optical H$\beta$ line (adapted from Turner et al. 1999). 
A weak, positive correlation is seen in the first case, and 
a very strong anticorrelation is observed in the latter. 
\label{fig:var_spec}
}
\end{figure}

This apparent scaling of PDS shape with luminosity appears to break
down at the lowest luminosities, however, with the least powerful AGN
showing less pronounced variability at least on short time scales
(e.g. Ptak et al 1998). This may indicate that this type of AGN has a
relatively large mass (e.g. Iyomoto \& Makishima 2000), but a
different emission mechanism and accretion mode, such as a radiatively
inefficient flow (e.g. Narayan \& Yi 1994; Blandford \& Begelman
1999). If there is a universal PDS scaling with mass, rather than
luminosity, then it is in principle possible to determine the mass
from the variability properties by scaling from Galactic black hole
candidates (GBHC) such as Cyg X-1 (e.g. Hayashida et al. 1998).
Although much recent progress has been made, we will need far better
data on the PDS of AGN before such estimates can be considered robust.
Uncertainties in our knowledge of the mass of Cyg X-1, and variability
of its noise characteristics (Belloni \& Hasinger 1990) introduce
further limitations to such methods.

No consensus has emerged as to the physical interpretation of either
the low frequency knee or the high frequency break. These features are
presumably related either to a physical size in the system or to some
characteristic time scale in the accretion disk. Examples of models
that have attempted to describe the PDS shape are those with extended,
inhomogeneous coronae (Kazanas, Hua \& Titarchuk 1997; Hua et
al. 1997), magnetic flares (Poutanen \& Fabian 1999) and
self-organized criticality (Mineshige et al. 1994).

\subsection*{Periodicities}

Mass-measurement through variability would be considerably eased if
there were precise periodicities that could be compared between
objects. Unfortunately such periodicities have been very difficult to
pin down. The analogy with GBHC suggests that AGN might well exhibit
quasi-periodic oscillations (QPOs), and there were some reports of
periodic behavior from EXOSAT and Ginga (e.g. Papadakis \& Lawrence
1993, 1995), with the most famous case being NGC 6814 (Mittaz \&
Branduardi-Raymont 1989; Done et al. 1992).  One case has been
disputed (Tagliaferri et al. 1996) and the the periodicity in NGC 6814
has been refuted decisively (Madejski et al. 1993). There are
remaining claims in NGC 4051 (Papadakis \& Lawrence 1995), RX
J0437.4-4711 ( Halpern \& Marshall 1996) and IRAS 18325-5926 (Iwasawa
et al. 1998). Given the difficulties in analysis of AGN power spectra
-- mainly associated with the severe distortion of most PDS by the
window function -- and the controversy surrounding the claims of
periodicities, many workers have been reluctant to accept the
contention that AGN exhibit any periodic behavior. Further data will
resolve this issue, which remains open.

\begin{figure}
\begin{center}
\includegraphics[height=200mm]{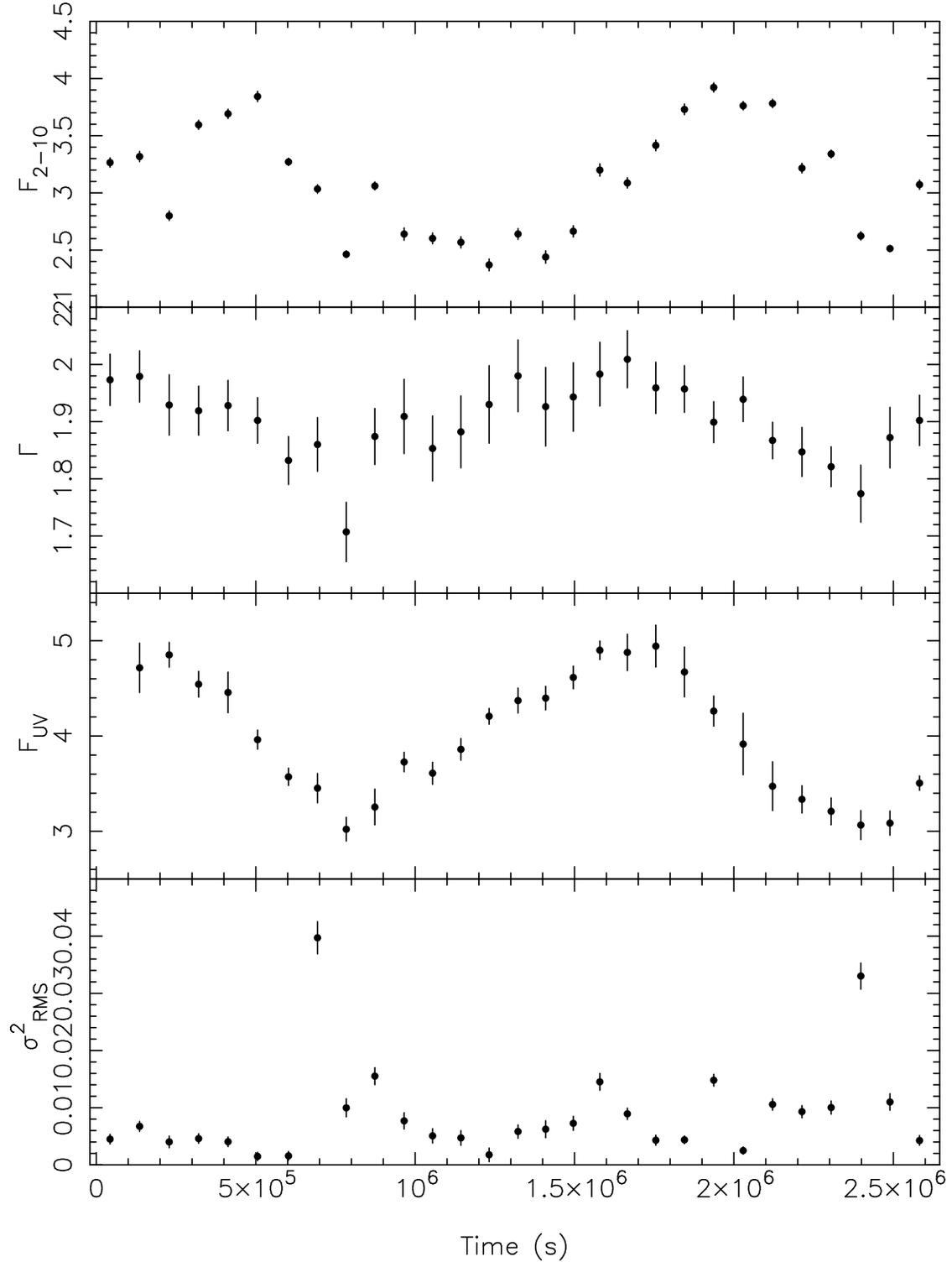}
\end{center}
\caption{Light curves for NGC 7469 of (top to bottom) the
2-10 keV X-ray flux ($F_{\rm 2-10}$), the X-ray spectral index ($\Gamma$), the
ultraviolet 1315\AA\ flux ($F_{\rm UV}$) and the fractional excess
variance computed on time scales of 1 day ($\sigma^{2}_{\rm RMS}$).
The spectral index is not well correlated with the X--ray flux, but
follows the UV flux (see also Fig.~\ref{fig:7469_ucor}). This
is supportive of thermal Comptonization models. The excess variance shows
unusual behavior, with periods of significantly enhanced variation
apparently coinciding -- or perhaps preceding by $\sim 1$d --
the times where the X-ray spectrum is hardest and the UV flux weakest.
The reason for this behavior is unclear, but it may represent some kind
of instability in the system.
\label{fig:rms_lc}
}
\end{figure}

\section*{CORRELATIONS WITH SPECTRAL PROPERTIES}

Green et al. (1993) first showed that the PDS of Seyferts depends on
the spectral shape, in that sources with steeper power-law slopes
appeared to be more variable on short time scales. This results has
now been confirmed by ROSAT and ASCA studies (Fig.~\ref{fig:var_spec};
Koenig, Staubert \& Wilms 1996; Fiore et al. 1998; Turner et al. 1999).
Green et al.'s original explanation of this behavior was that the
harder sources have a more dominant contribution from ``Compton
reflection'', although this seems unlikely given the relatively small
contribution that reflection makes in the $\sim 2-6$~keV band to which
the EXOSAT data were most sensitive.

\begin{figure}[tb]
\begin{center}
\includegraphics[width=180mm]{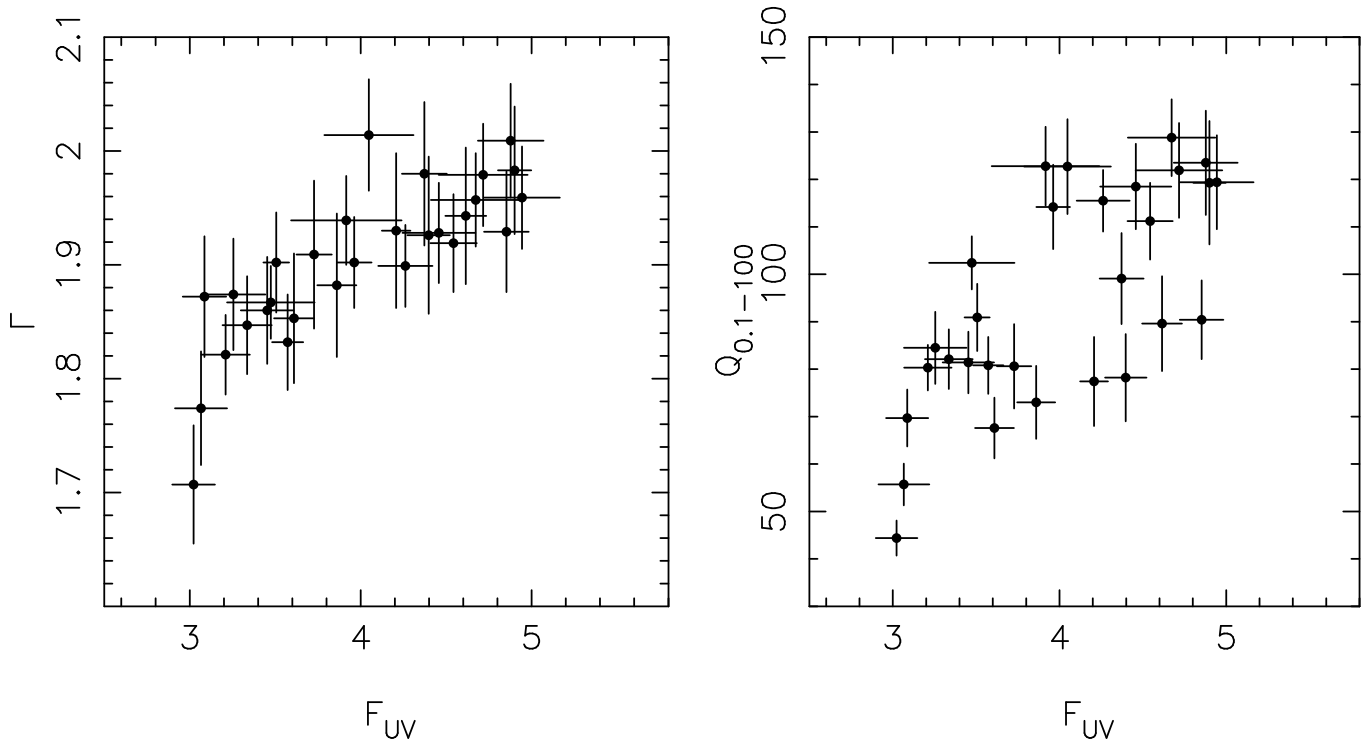}
\end{center}
\caption{The left panel shows the correlations between the UV flux of 
NGC 7469 at 1315\AA\ and
X-ray spectral index ($\Gamma$). There is a strong
correlation, as expected in Comptonization
models if the UV represents the seed photons. Increased seed
flux would cool the X-ray Comptonizing corona, resulting in a steeper
slope. The right panel shows the relationship between the UV flux 
and the broad-band (0.1-100 keV) X-ray photon flux. This strong
correlation again supports
Comptonization, as we expect in such models that one scattered UV
photon would ultimately emerge as an X-ray, although we note that
a narrow band might not show such a correlation 
(c.f. Fig.~\ref{fig:rms_lc}).
\label{fig:7469_ucor}}
\end{figure}

New impetus for the study of this correlation has arisen from the
realization that many of the properties of active galaxies are
inter-related. Boroson \& Green (1992) showed numerous optical line
widths and strengths showed correlations, most notably $H\beta$ width
and Fe {\sc ii} strength, and identified a driving parameter
``eigenvector 1'' through principal component analysis. Boller et
al. (1996) brought the X-rays into this picture, by showing that the
soft X-ray spectral index of AGN is also correlated with H$\beta$
width. These observations have concentrated much attention on
the the subclass of AGN known as ``Narrow-Line Seyfert 1'' (NLS1)
galaxies, which exhibit the narrowest permitted (specifically
H$\beta$) lines, strongest Fe {\sc ii} emission and steepest X-ray
slopes. Some NLS1 were also shown to have dramatic X-ray variability
properties in ROSAT observations (e.g. Boller et al. 1997), fitting
in with the correlation of X-ray slope with variability amplitude
just described. 

Turner et al. (1999) demonstrated explicitly the correlation between
X-ray variability amplitude and FWHM H$\beta$
(Fig.~\ref{fig:var_spec}).  Leighly (1999) has shown similar results,
by showing that the NLS1 occupy a particular part of the
$\sigma^{2}_{\rm RMS}$ vs.  luminosity diagram (also demonstrated in
Fig.~\ref{fig:var_lum}).  These observations show that X-ray
variability amplitude can be added to the slew of correlated
properties related to ``eigenvector 1''. This is of clear importance,
as X-ray variability must be related to some fundamental
energy-generation process. The current wisdom is that that NLS1 have
lower masses and higher accretion rates than objects with broader
optical lines (Pounds et al. 1995). The higher variability amplitude
in NLS1 fits into this picture, as they would be smaller than
broad-line objects at the same luminosity. Many of the details are
unclear, however, given that the origin of the variability - and even
of the X-rays themselves, is not very well known. For instance, the
steeper slopes observed in NLS1 clearly show that their Comptonizing
coronae have different physical characteristics.  This could result in
differing variability characteristics, even at the same mass.

\section{NGC 7469: A variability case study}

\begin{figure}[tb]
\begin{center}
\includegraphics[width=120mm]{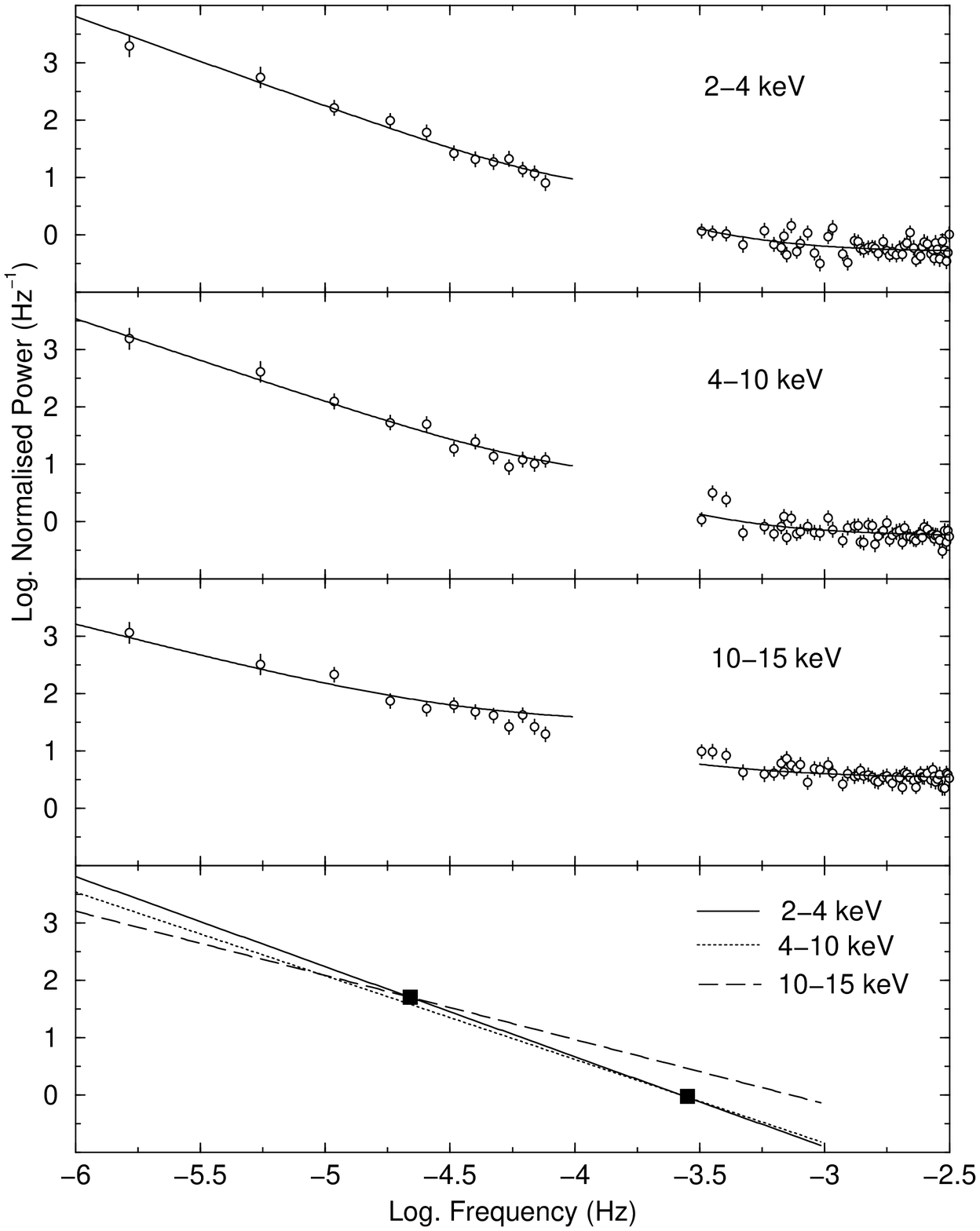}
\end{center}
\caption{PDS of NGC 7469 in three different energy bands (Nandra \&
Papadakis 2001). These were constructed in a two stage process, using
the orbitally-binned data for the low frequency part, and an average
of the intra-orbit 16s binned data for the high frequencies. The
bottom panel shows the best-fit models to the PDS. These show that at
low frequencies, more variability power is present in the soft X-ray
PDS, which is is consistent with the long-term variations being due to
changes in the Compton seed.  The PDS is {\it flatter} at higher
energies, however, in contrast to the simplest expectations of
Comptonization.  On time scales shorter than about 1d, there is
relatively more variability at high energies, implying another
variability process, presumably related to the coronal heating
mechanism, dominates the most rapid variations.
\label{fig:7469_pds}}
\end{figure}

RXTE undertook a long ($\sim 30$~d), quasi-continuous monitoring
observation of the Seyfert 1 galaxy NGC 7469, simultaneously with IUE
(Nandra et al. 1998, 2000). RXTE showed strong variations in the X-ray
flux. The UV flux also varied (Wanders et al. 1997), but the most
rapid changes seen in the X-rays were not present in the UV light
curve. The fluxes in the 2-10 keV band and 1315\AA\ bands were not
well correlated, which contradicts naive expectations from the Compton
upscattering models that identify the UV as the seed photons. The
lack of a clear relationship is also counter to simple
``reprocessing'' models for the UV emission, which will not be
discussed in detail here, but have many attractive features.

The apparent discrepancy between the data and these model is explained
when the \xte \ spectral data are taken into account
(Fig.~\ref{fig:rms_lc}).  This shows that, although the 2-10 keV
and UV fluxes are not well correlated, there is a very strong
relationship between the UV flux and the X-ray spectral index
(measured in the 2-20 keV band; Fig.~\ref{fig:7469_ucor}). This is
expected in Comptonization models: as the UV seed flux increases so
does the cooling, resulting in a softer spectrum (e.g. Haardt,
Maraschi \& Ghisselini 1997; Zdziarski, Lubinski \& Smith 1999).  The
lack of correlation between the UV seed photons and the 2-10 keV X-ray
flux may then be explained by a bandpass effect, with the 2-10 keV
flux being strongly affected by the X-ray spectral
variations. Extrapolation of the X-ray power law over a wider band
shows that both the soft X-ray/EUV flux and the broad-band X-ray
photon flux (Fig.~\ref{fig:7469_ucor}) are both well correlated with
the UV emission. The former is helpful for the reprocessing
models. The latter is indeed what might be expected in the
Comptonization scenario, in the sense that the correlation should be
between the photons, and not necessarily the flux. An increase in the
number of seed photons results in a corresponding increase in the
number of X-ray photons, but the total energy output in a relatively
narrow spectral band also depends on other factors, such as the
optical depth, temperature and energy input into the corona. The NGC
7469 spectral data therefore offer strong support for the
Comptonization model. Correlations have been observed between the UV
and EUV emission in at least one Seyfert (Marshall et al. 1997) and
between the EUV and the X-ray emission (Uttley et al. 2000). The
latter may even show a time delay in the sense that the EUV emission
leads (Chiang et al. 2000), further supporting the Comptonization
hypothesis. Finally, a long BeppoSAX observation of NGC 5548 showed an
explicit connection between the spectral slope in the X-ray band and
the high energy cutoff, which indicates the temperature of the corona
(Petrucci et al. 2000). During a ``flare'' in the X-ray emission, the
X-ray spectrum was seen to steepen and the temperature to reduce,
exactly as expected from the thermal Comptonization models if the UV
also increased during the flare. Unfortunately no UV data were
available at the time of the BeppoSAX observation, but combining the
results with those obtained for NGC 7469 leads to compelling evidence
that thermal Comptonization is indeed the dominant radiation mechanism
producing X-rays in radio quiet AGN.

The fact that the X-rays show much more rapid variations than the UV
is, on the other hand, inconsistent with a simple picture in which
energy is dissipated in the Comptonizing corona at a constant rate,
and variations in the UV seed photons cause both the X-ray variability
and cooling of the corona. In this case, as the X-rays are a secondary
component, they should vary less rapidly than the seed photons, or
with a lower amplitude on a fixed time scale.  There are at least two
ways around this that preserve the Comptonization model. First, the UV
may not be the seed photon population, but may merely be correlated
with the true seed, presumably the EUV coming from smaller radii. This
could then be more rapidly variable than both the X-rays and the UV,
but have the variations more smoothed in the UV than the
X-rays. Second, the very rapid X-ray variations might arise from a
process unrelated to the Compton seed, most obviously the coronal
heating mechanism.

The PDS appears to resolve this ambiguity (Nandra \& Papadakis 2001;
Fig.~\ref{fig:7469_pds}).  There is no obvious low frequency knee down
to a frequency of about $10^{-6}$~Hz, which is not surprising given
the luminosity is rather similar to NGC 3516 (Fig.~\ref{fig:pds}),
where the knee is an order of magnitude lower in frequency. These time
scales were not sampled by the NGC 7469
campaign. Fig.~\ref{fig:7469_pds} shows the PDS obtained at three
different energies. There are no obvious features in any, although at
high frequency the PDS do show some structure that is not well fit by
the power-law model. The most striking thing about
Fig.~\ref{fig:7469_pds}, however, is the clear difference in the power
spectra as a function of energy. At low frequencies there is clearly
more power in the soft X-ray PDS, but the opposite is true at high
frequencies. This is parameterized in the bottoms panel of the Figure
by a power law that is significantly flatter at higher
energies. Although the signal-to-noise ratio in the 10-15 keV band is
lower than the softer energy bands, making it more susceptible to
errors in the background subtraction, these are unlikely to cause the
observed effect. A similar analysis of ``blank sky'' background
observations reveals no such effect, nor any excess of power at high
frequencies which might bias our results.  It is interesting to note
that a similar behavior ``hardening'' of the PDS has been reported for
Cyg X-1 (Nowak et al. 1999) and is very much opposite to that expected
from Comptonization (e.g. Hua \& Titarchuk 1996) where the PDS should
be steeper at high energies.  High energy photons undergo more
scatterings, ``washing out'' the high frequency variability.

The implication is that the most rapid variations are not caused by
variations in the Compton seed - regardless of whether the seed
photons are UV or EUV - but arise from a separate mechanism. We
identify this with the process that heats the X-ray corona (for
example, the magnetic flares of Poutanen \& Fabian, 1999, although
they predict the high frequency power spectrum should be similar at
all energies). Assuming that the X-rays are indeed produced by Compton
upscattering, it seems highly unlikely that this ``hardening'' can be
produced in a model where the X-rays are produced in a single,
coherent region. In such a scenario, changes in the dissipation rate
would presumably affect all energies in the same way. Our data require
a patchy corona, or localized ``flares''. One possibility is that the
flares in the inner regions are typically hotter. They would therefore
account for more of the high energy photons, but be more rapidly
variable because of their smaller size scale. We await a detailed
model that explains this flattening of the PDS with increasing photon
energy, which appears to be relevant to both GBHC and, now, AGN.

A final intriguing result from the NGC 7469 campaign is illustrated by
the bottom panel of Fig.~\ref{fig:rms_lc}. This shows the
fractional RMS variability observed in the $\sim 1 d$ integrations
used to derive the spectral parameters. The variability amplitude
clearly changes from day to day. Most dramatically, it seems to show a
strong increase at the times (or perhaps just preceding the times)
when the UV flux is weakest and the X-ray spectrum the hardest. The
behavior of this object thus contrasts with that when comparing
objects (Fig~\ref{fig:var_spec};sources with flat spectra are {\it
less} variable on $\sim d$ time scales). The reason for this is not
clear at this point, but it may be related to some form of instability
in the corona. For example, as the spectrum hardens it may reach a
critical point at which electron-positron pair production becomes
important, inducing additional variability.

\section*{SUMMARY}

Variability studies of AGN have been somewhat neglected since the {\it
EXOSAT} era, but new data from \xte\ and interesting results from the
GBHC observations is providing new impetus. It appears that
AGN variability shows great similarity to that of GBHC. 
Indeed, QPOs are the only important characteristic of Galactic
binary variability that has not yet been established firmly in
AGN. Population studies have shown X-ray variability to be strongly
related to several other properties. The luminosity dependence - which
may extend all the way down to GBHC - is indicative of a ``universal''
power spectrum that scales approximately with the length scale of the
source. Much more data is required before this hypothesis can be
firmly established, however. There is also a strong dependence of
X-ray variability on spectral characteristics, in the sense that AGN
with softer spectra - notably the subclass of NLS1 - are more
variable. This remains unexplained, but clearly indicates that
``Eigenvector 1'' is related to something fundamental about the energy
generation process. Detailed variability studies of some individual
objects lends strong support to the idea that the emission mechanism
for the X-rays is thermal Comptonization. In one source, NGC 7469,
however, there is clear evidence that there is an additional
variability mechanism on top of changes in the Compton seed. This must
be related to the process that heats the Comptonizing corona, and
apparently affects the hard X-rays more than the soft. Another
intriguing phenomenon in this source is the observation of a dramatic
increase in the X-ray variability amplitude at the time when the
spectrum is hardest - in contrast to the behavior observed when
comparing objects. This may be related to some kind of instability,
perhaps involving electron-positron pair production.  Whatever the
origin of these phenomena, it is clear that variability still has much
to tell us about both the radiation mechanism and the dissipation
process that causes active galaxies to emit X-rays.

\section*{ACKNOWLEDGEMENTS}

I am most grateful to my collaborators on the NGC 7469 project: Jean
Clavel, Rick Edelson, Ian George, Matt Malkan, Richard Mushotzky, Brad
Peterson and Jane Turner for their efforts in bringing the observation
to fruition. This work also relied heavily on the expertise of Iossif
Papadakis, of the University of Crete. I acknowledge the financial the
support of NASA, via grant NAG5-7067 to the Universities Space
Research Association.

\end{document}